\documentclass[aps,prb,twocolumn,groupedaddress,showpacs,floatfix]{revtex4}
\usepackage{graphicx}
\begin{document}
\pagenumbering{arabic} 
\pagestyle{plain}
\title{$^{1}$H-NMR spin-echo measurements of the spin dynamic properties in $\rm{\lambda}$-(BETS)$_{2}$FeCl$_{4}$}
\author{Guoqing Wu,$^{1}$ P. Ranin,$^{1}$ G. Gaidos,$^{1}$ W. G. Clark,$^{1}$ S. E. Brown,$^{1}$ L. Balicas,$^{2}$ and L. K. Montgomery$^{3}$}
\affiliation{$^{1}$Department of Physics and Astronomy, UCLA, Los Angeles, California 90095-1547, USA}
\affiliation{$^{2}$National High Magnetic Field Laboratory, Florida State University, Tallahassee, Florida 32306, USA}
\affiliation{$^{3}$Department of Chemistry, Indiana University, Bloomington, Indiana 47405, USA}
\date{\today}
\begin{abstract}
    $^{1}$H-NMR spin-echo measurements of the spin-echo decay $M(2\tau)$ with a decay rate 1/$T_{2}$ under applied magnetic field $\mathbf{B}$$_{0}$ = 9 T along the $a$ axis over the temperature ($T$) range 2.0$-$180 K are reported for a single crystal of the organic conductor $\rm{\lambda}$-(BETS)$_{2}$FeCl$_{4}$. It provides the spin dynamic properties in the paramagnetic metal (PM) and antiferromagnetic insulator (AFI) states as well as across the PM-AFI phase transition. A large slow beat structure in the spin-echo decay is observed with a typical beat frequency of $f_B$ $\sim$ 7 kHz that varies across the spectrum. Its origin is attributed to the interactions between protons that are very strongly detuned by the large inhomogeneous field on a microscopic distance scale that is generated by the Fe$^{3+}$ moments (spin $S_d$ = 5/2). A simple phenomenological model provides an excellent fit to the data. The values of 1/$T_{2}$ in the PM phase are consistent with a $T-$independent contribution from the proton-proton dipole interaction plus the proton spin-lattice relaxation rate (1/$T_{1}$) [W. G. Clark $et ~al.$, Appl. Magn. Reson. $\bf{27}$, 279 (2004)], which has a significant contribution only above $\sim$ 20 K. At the PM-AFI transition (3.5 K), there is a discontinuous drop in 1/$T_{2}$ by $\sim$ 34 $\%$, indicating that the transition is first order, consistent with prior reports. Two possible main contributions to this drop are discussed. They are based upon the change in the local magnetic field caused by the change in the orientation of the Fe$^{3+}$ moments at the transition.    
\end{abstract}
\maketitle
\section{Introduction}
     Organic conductors have been of considerable interest because of their characteristic properties in low-dimensional electronic system.\cite{ishiguro, clark1, gr} The quasi-two dimensional (2D) triclinic (space group P$\bar{1}$) salt, $\rm{\lambda}$-(BETS)$_{2}$FeCl$_{4}$, where BETS is bis(ethylenedithio)tetraselenafulvalene (C$_{10}$S$_{4}$Se$_{4}$H$_{8}$), has been one of the most attractive materials among various organic conductors.\cite{uji, kobayashi1, tokumoto, brossard, akutsu1} Below a magnetic field ($\mathbf{B}_{0}$) of about 14 T, as the temperature ($T$) is lowered, it has a transition from a paramagnetic metal (PM) to an antiferromagnetic insulating (AFI) phase, and at higher fields and low $T$ there is a PM to a field-induced superconducting (FISC) phase.\cite{uji, kobayashi1, tokumoto, brossard} Although it has been suggested\cite{uji, akutsu1, balicas} that the FISC phase can be explained by the Jaccarino-Peter compensation effect,\cite{jaccarino} one problem with this interpretation is that the FISC state in $\lambda$-(BETS)$_{2}$FeCl$_{4}$ could be destroyed by a very small amount of out-of-plane magnetic field.\cite{uji,balicas} 

    An important aspect of the PM to AFI transition in $\lambda$-(BETS)$_{2}$FeCl$_{4}$, which is one of the main topics of this paper, is that the metal-insulator (M-I) transition and that of the paramagnetic to antiferromagnetic order occur at the same $T$, i.e., $T_{N}$ = $T_{\rm{MI}}$, ($T_{N}$ is the N$\acute{\rm{e}}$el temperature and $T_{\rm{MI}}$ is the M-I transition). This coordination of properties is confirmed by electrical resistivity, heat capacity, and magnetic susceptibility measurements where a high spin $S_d$ = 5/2 state for the Fe$^{3+}$ is observed.\cite{brossard, kobayashi2} It has been stated that in this material the AF ordering of the Fe$^{3+}$ ion spins is mediated by the $\pi$-electrons, while the direct magnetic dipole-dipole interaction between the Fe$^{3+}$ $-$ Fe$^{3+}$ spins is negligible because the neighboring Fe$^{3+}$ are more than $\sim$ 6 $\rm{\AA}$ apart.\cite{kobayashi1, kobayashi3} It should be noted that the phase diagram of the iso-structural nonmagnetic and non-3$d$-electron analog $\rm{\lambda}$-(BETS)$_{2}$GaCl$_{4}$ is completely different from that of $\rm{\lambda}$-(BETS)$_{2}$FeCl$_{4}$.\cite{kobayashi4} So far, the role of the $\rm{\pi}$-d interaction for the PM-AFI phase transition has not been investigated with direct experimental evidence. 

    In this paper, we report proton nuclear magnetic resonance ($^{1}$H-NMR) spin-echo measurements in an applied magnetic field $\bf{B}_{0}$ = 9 T aligned along the $a$-axis for 2.0 K $\leq$ $T\leq$ 180 K on a single crystal of the organic conductor $\rm{\lambda}$-(BETS)$_{2}$FeCl$_{4}$. They include the spin-echo amplitude $[M(2\tau)]$ as a function of the spin-echo arrival time (2$\tau$) that provides the spin-echo decay rate 1/$T_{2}$. These measurements probe the dynamic properties of the nuclear and electron spins in the PM and AFI states as well as those across the PM-AFI phase transition. Also, the electrical resistivity ($\rho$) of $\rm{\lambda}$-(BETS)$_{2}$FeCl$_{4}$ as a function of $T$ at several values of $\bf{B}_{0}$ $\parallel$ $a$ on the same sample is reported.

    One important result of this investigation is that $M(2\tau)$ has large, slow beats that are explained by the proton$-$proton dipole interaction. Also, it is shown that above $\sim$ 20 K a significant contribution to $1/T_{2}$ is the the proton spin-lattice relaxation rate ($1/T_{1}$), which is dominated by the dipolar field fluctuations from the 3d Fe$^{3+}$ ions. There is also a clear discontinuity in 1/$T_{2}$ across the PM-AFI phase transition, which indicates a significant change in the slow fluctuations of the local magnetic field at the $^{1}$H-sites on traversing the PM to AFI phase transition.

     The rest of this paper is organized as follows. Section II presents the experimental details and Sec. III has the experimental results for $M(2\tau)$, a fit to $M(2\tau)$ using a phenomenological model, and the data of 1/$T_{2}$. Section IV discusses in more detail the origin of the slow beats and the mechanism of 1/$T_{2}$. The conclusions are stated in Sec. V. 
\section{experimental details}
     Single crystal $\rm{\lambda}$-(BETS)$_{2}$FeCl$_{4}$ samples have a needle-shape. They were prepared as described by Montgomery et al with a standard electrochemical oxidation method.\cite{montgomery} The dimensions of the sample used for these measurements are (1.2 $\pm$ 0.1) mm $\times$ (0.065 $\pm$ 0.010) mm $\times$ (0.018 $\pm$ 0.005) mm, which corresponds to (3.8 $\pm$ 1.8) $\mu$g in mass and (2.7 $\pm$ 1.3) $\times$ 10$^{16}$ protons. 

     To obtain a filling factor that provides a viable sensitivity for a single crystal, a very small NMR coil was used. It was made using 40 turns of 0.025 mm diameter bare copper wire wound on a 0.075 mm diameter wire form. The single crystal sample was oriented in the coil with $\bf{B}_{0}\parallel a$, where $a$ is defined as $\perp$ $c$ in the plane of the largest face. An estimated uncertainty in the angle setting is $\sim$ $\pm$ 5$^{\circ}$. The proton NMR spin-echo measurements were made using a spectrometer and probe built at UCLA and a commercial 9 T magnet. The typical NMR frequency ($\nu$) used was 383.0 MHz (the Larmor frequency $\nu_{0}$ = $\gamma_{p}B_0$ = 382.6935 MHz, where $\gamma_{p}$ = 42.5759 MHz/T is the proton gyromagnetic ratio, and $B_0$ = 8.9885 T). For simplicity, the value of $B_{0}$ is often referred to as 9 T.

     Measurements of $M(2\tau)$ were done using standard spin-echo methods in which a $\pi$/2 pulse $p_{1}$ (typical value of $p_{1}$ = 0.2 $\mu$s) is used for preparation, followed a time $\tau$ later ($\tau$ varies from 5 $\mu$s up to 500 $\mu$s) by a second pulse $p_{2}$ whose angle $\beta$ is adjustable to maximize the spin-echo magnitude (typically $\beta$ = 3$\pi$/4 here). Each echo transient was recorded in the time domain with a frequency response width of about $\pm$ 0.4 MHz.

     As discussed in Section III, 1/$T_{2}$ and the slow beat frequency $f_{B}$ were obtained using a phenomenological model to fit the $M(2\tau)$ vs $2\tau$ data. The $^{1}$H-NMR spectrum used to identify the spectral peak and for the corresponding 1/$T_{2}$ measurements was obtained by measuring the frequency-swept spectra with frequency range covering from 370 up to 400 MHz. The spectra were analyzed with frequency-shifted and summed Fourier transform processing\cite{clark3} of the spin-echo. More complete details of the NMR measurements are presented elsewhere.\cite{wu-lbetsp3, clark4}

     The electrical resistivity measurements as a function of $T$ with $\mathbf{B}_{0}$ $\parallel$ $a$ axis were made using a standard four probe technique, with 13 $\mu$m diameter Au-wires and Ag paint for contacts. 
\section{Experimental results}
\subsection{Electrical resistivity}
\begin{figure}
\includegraphics[scale=0.36]{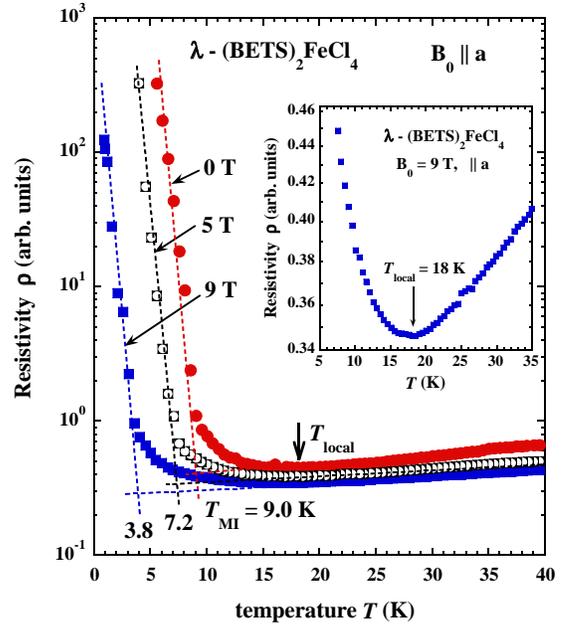}
\caption{(Color online) Electrical resistivity $\rho$ (arb. units) as a function of $T$ for the single crystal of $\rm{\lambda}$-(BETS)$_{2} $FeCl$_{4}$ with $\bf{B}_{0}~||~a$ for $B_{0}$ = 0, 5, and 9 T. The dashed lines are for the determination of $T_{\rm{MI}}$ at the corresponding $B_{0}$. The thick solid arrow indicates $T_{\rm{local}}$, where $\rho$ has a minimum. The inset shows $\rho$ for 5 K $\leq T \leq$ 35 K at $B_{0}$ = 9 T. \label{fig1}}
\end{figure}
     Figure 1 shows the electrical resistivity $\rho$ (normalized relative to the room temperature value) of $\rm{\lambda}$-(BETS)$_{2}$FeCl$_{4}$ under corresponding applied magnetic field $B_{0}$ = 0, 5, and 9 T at $\bf{B}_{0}$ $\parallel$ $a$. This alignment of $\bf{B}_{0}$ is the same as that used for the NMR measurements. There is a broad minimum in $\rho$ around $T$ $\sim$ $T_{\mathrm{local}}$ = 18 K (solid arrow), where the $\rm{\pi}$-electrons begin to be localized\cite{akutsu1, akutsu2, akutsu3} (shown in more detail in the inset). It is believed that this minimum is caused by the $\rm{\pi}$-electron localization, which is related to the strong inter-BETS-molecular and $\rm{\pi}$-d interactions.\cite{akutsu1, akutsu2, akutsu3, tanaka}

     Also, increasing $B_{0}$ reduces $T_{\rm{MI}}$ to lower $T$, with $T_{\rm{MI}}$ = 9.0, 7.2 and 3.8 K for $B_{0}$ = 0, 5 and 9 T, respectively. The value of $T_{\mathrm{local}}$ is essentially independent of $B_{0}$ within the experimental uncertainty. The values of $T_{\rm{MI}}$ agree with those of S. Uji et al.\cite{uji} and $T_{\rm{MI}}$ for $\bf{B}_{0}$ $\parallel$ $a$ is slightly ($\sim$ 0.5 K) larger than that for $\mathbf{B}_{0}$ $\parallel$ $c$. 
\subsection{Slow beats in $M$(2$\tau$) and their fit parameters}
\begin{figure}
\includegraphics[scale= 0.35]{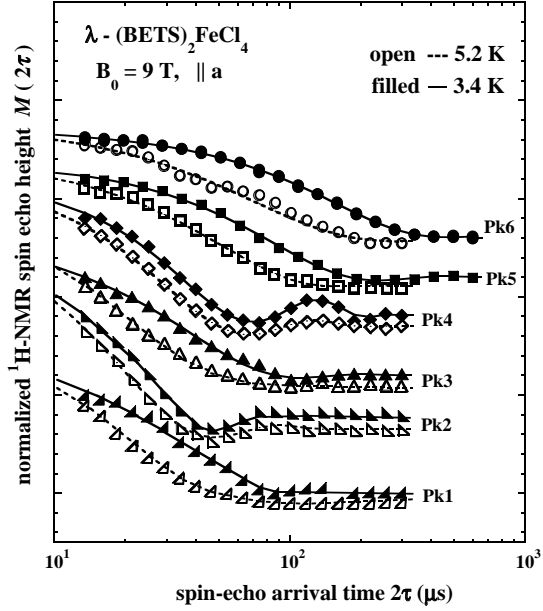}
\caption{Proton $M$(2$\tau$) as a function of 2$\tau$ for different spectral peaks at 3.4 K (solid symbols) and 5.2 K (open symbols) for the single crystal of $\rm{\lambda}$-(BETS)$_{2} $FeCl$_{4}$ with $\bf{B}_{0}$ = 8.9885 T $||~a$. The solid (3.4 K) and dashed (5.2 K) lines are the fit to Eq. (1) and the peaks are identified in Fig. 3. \label{fig2}}
\end{figure}

      Figure 2 plots the typical $^{1}$H-NMR spin-echo height $M$(2$\tau$) as a function of spin echo arrival time 2$\tau$ at $T$ = 3.4 K and 5.2 K with $\mathbf{B}_{0}$ = 9 T $||$ $a$ for the six major peaks in the NMR spectrum.
\begin{figure}
\includegraphics[scale= 0.35]{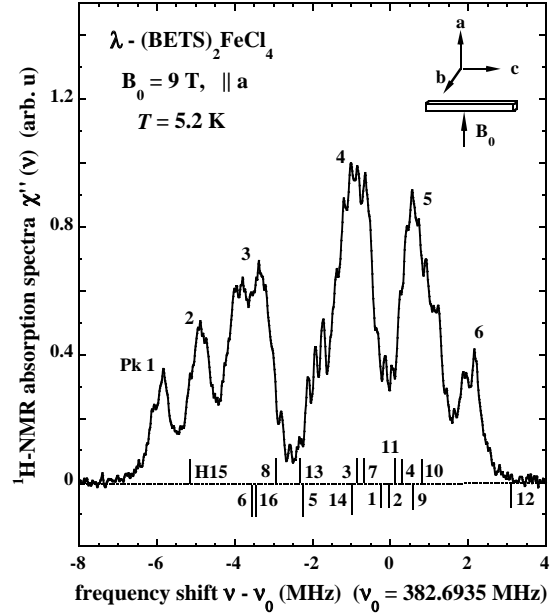}
\caption{Proton absorption spectrum $\chi^{''}$ ($\nu$) as a function of $\Delta\nu=\nu-\nu_{0}$ for $\nu_{0}$ = 382.6935 MHz at $T$ = 5.2 K for the single crystal of $\rm{\lambda}$-(BETS)$_{2} $FeCl$_{4}$ with $\bf{B}_{0}$ = 8.9885 T $\parallel$ $a$. The number at the top of each peak is its index number and the vertical lines at the bottom are the calculated $\Delta\nu$ for each of the 16 proton sites. \label{fig3}}
\end{figure}
\begin{figure}
\includegraphics[scale= 0.37]{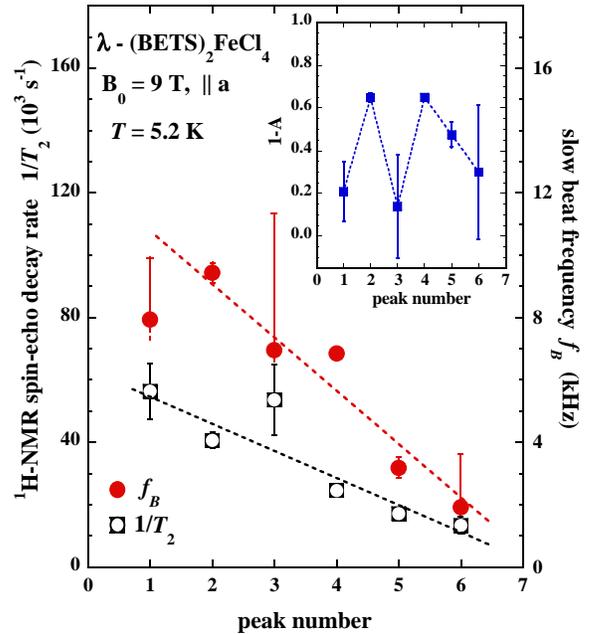}
\caption{(Color online) $\Pr$oton 1/$T_{2}$, $f_{B}$, and $(1-A)$ (inset) as a function of the peak index (Fig. 3) at 5.2 K for the single crystal of $\rm{\lambda}$-(BETS)$_{2} $FeCl$_{4}$ with $\bf{B}_{0}$ = 8.9885 T $||~a$. \label{fig4}}
\end{figure}
In Fig. 3, the full proton NMR absorption spectrum $\chi^{''}(\nu)$ is plotted as a function of $\Delta \nu = \nu - \nu_{0}$ with the peak label numbers at the top of the peak. Along the bottom, the vertical lines show $\Delta\nu$ at each of the 16 proton sites obtained from the calculated component of the Fe$^{3+}$ dipole field parallel to $\mathbf{B}_{0}$ at the proton sites.\cite{wu-lbetsp3} Because the model used to calculate $\Delta\nu$ does not include all Fe$^{3+}$ interactions, the assignment of proton sites to peak numbers has a substantial uncertainty that is difficult to quantify. 
\begin{figure}
\includegraphics[scale= 0.33]{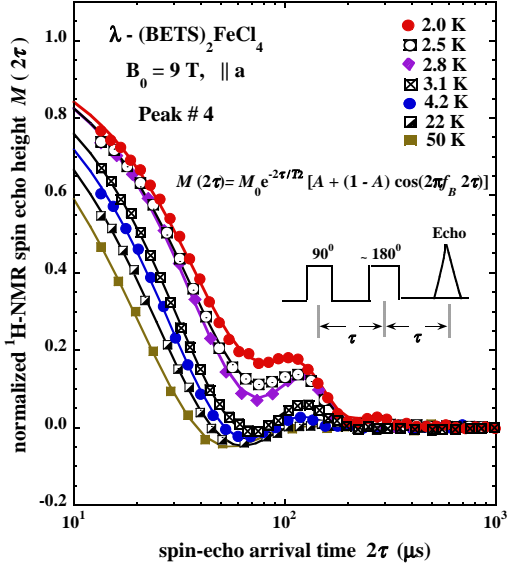}
\caption{(Color online) Proton spin-echo height $M$(2$\tau$) of the spectral peak 4 as a function of 2$\tau$ and several $T$ for the single crystal of $\rm{\lambda}$-(BETS)$_{2} $FeCl$_{4}$ with $\bf{B}_{0}$ = 8.9885 T $||~a$. The solid lines are the fit obtained with Eq. (1). \label{fig5}}
\end{figure}
\begin{figure}
\includegraphics[scale= 0.33]{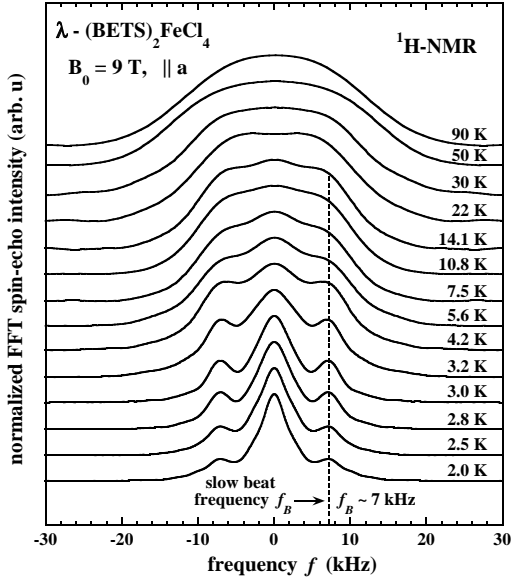}
\caption{Normalized FFT of $M(2\tau)$ for the spectral peak 4 for the single crystal of $\rm{\lambda}$-(BETS)$_{2} $FeCl$_{4}$ with $\bf{B}_{0}$ = 8.9885 T $||~a$. The dashed line indicates $f_{B}$ $\sim$ 7 kHz. \label{fig6}}
\end{figure}
\begin{figure}
\includegraphics[scale= 0.33]{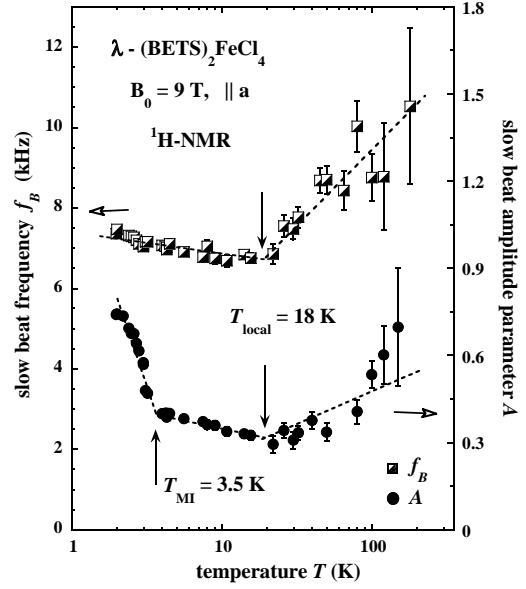}
\caption{$T-$dependence of $f_B$ and $A$ in Eq. (1) as a function of $T$ for the spectral peak 4 of the single crystal of $\rm{\lambda}$-(BETS)$_{2} $FeCl$_{4}$ with $\bf{B}_{0}$ = 8.9885 T $||~a$. The dashed lines are a guide to the eye. \label{fig7}}
\end{figure}

     An important property of $M$(2$\tau$) is that its shape depends on both $T$ and which spectral peak is measured.

     The main features of $M$(2$\tau$) are: (1) an overall exponential decay rate 1/$T_{2}$ and (2) for some of the spectral lines a slow oscillation with the beat frequency $f_{B}$. A phenonomenological fit of each line is obtained using
\begin{equation}
M(2\tau)=M_{0}[A+(1-A)\cos(2\pi f_{B}2\tau)]\exp(-\frac{2\tau}{T_{2}}),\\
\end{equation}
where $T_{2}$, $f_{B}$, and $A$ are the fit parameters. $M(2\tau)$ is normalized to 1 at 2$\tau$ = 0. A similar formula was developed to model $M(2\tau)$ in a situation where $f_{B}$ was dominated by the magnetic dipolar interactions between nuclear spins. \cite{abragam, ansermet} As shown by the solid and dashed lines in Fig. 2, Eq. (1) provides an excellent fit to $M(2\tau)$ for all the spectral peaks in $\rm{\lambda}$-(BETS)$_{2}$FeCl$_{4}$.

     Figure 4 shows the plot of 1/$T_{2}$, $f_{B}$, and $(1-A)$ (inset) for each of the spectral peaks at 5.2 K. There is a substantial variation in all of them across the spectrum. An interpretation of this effect is given in Section IV. 

     Since the peak 4 has the largest signal and a well defined $f_{B}$, it was used to examine the properties of $M(2\tau)$ over a wide range of $T$ (2$-$180 K). Figure 5 shows a plot of $M$(2$\tau)$ as a function of 2$\tau$ at the same $\mathbf{B}_{0}$ for a few $T$ with $\nu$ adjusted to follow the peak 4. The solid lines are the fit using Eq. (1). Two important features seen in Figs. 2$-$5 are that the shape of $M(2\tau)$ depends on both $T$ and the peak number.

     An alternative way to obtain $f_{B}$ is to apply the fast fourier transform (FFT) algorithm to $M$(2$\tau)$. This was done by mirroring $M$(2$\tau)$ into the $-2\tau$ domain, using a cubic spline interpolation to obtain equally spaced linear time intervals, and applying the FFT algorithm to this array. The results are shown in Fig. 6, where the normalized FFT of the amplitude of the spin-echo decay is plotted as a function of frequency $f$. 

     There, it is seen that at low $T$, there is a peak centered at zero frequency plus a low frequency peak shifted by $f_{B}$ $\sim$ 7 kHz. At higher $T$, these peaks are smeared out by the progressively wider distribution, which corresponds to the shortening of $T_{2}$. Even with this smearing, it is seen in Fig. 5 that the beats persist to relatively high $T$.

     An important property of $f_{B}$ is that it does not change across the PM-AFI phase transition. This behavior implies that the internal source that generates the beats is independent of this phase transition.

     The slow beat fit parameters $f_B$ and $A$ for the spectral peak 4, obtained using Eq. (1), are shown as a function of $T$ in Fig. 7. Our best estimate of the uncertainty in the parameters is approximately the size of the symbols below 10 K and the error bars at higher $T$. Figure 7 shows in more detail than Fig. 6 that $f_B$ = (7.0 $\pm$ 0.3) kHz below 18 K and that there is no obvious change in $f_B$ across the PM-AFI phase transition.

     On the other hand $A$ does increase from $A$ = 0.40 $\pm$ 0.05 at 4 K to 0.73 $\pm$ 0.02 at 2 K when $T$ is lowered through the PM-AFI transition. Also, above $\sim$ 18 K both $f_B$ and $A$ increase slightly with temperature. It is noticeable that the temperature at the minimum of $\rho$ (Fig. 1) also matches these features for the slow beat fit parameters. Although their physical origins are not fully understood, an interpretation of these features is presented in Sec. IV A. 
\subsection{$^{1}$H-NMR spin-echo decay rate 1/$T_{2}$}
     The upper curve in Fig. 8 shows 1/$T_{2}$ as a function of $T$ for $\bf{B}_{0}$ = 9 T $\parallel a$ obtained from the fits using Eq. (1). There, a sudden drop of 1/$T_{2}$ from (23.3 $\pm$ 1.2) $\times$ 10$^{3}$ s$^{-1}$ at 4.0 K in the PM phase to (15.5 $\pm$ 0.5) $\times$ 10$^{3}$ s$^{-1}$ at 3.0 K in the AFI phase is seen across $\sim$ 3.5 K. Below $\sim$ 3.5 K, 1/$T_{2}$ is independent of $T$. Near 50 K, there is a peak in 1/$T_{2}$. Also shown for consideration later are the $^{1}$H-NMR spin-lattice relaxation rate 1/$T_{1}$ (lower curve) and 1/$T_{1}$ $-$ 1/$T_{2}$ (middle curve) vs $T$ at the same field and orientation for the same $\rm{\lambda}$-(BETS)$_{2} $FeCl$_{4}$ sample. Measurements using increasing and decreasing values of $T$ do not show hysteresis in 1/$T_{2}$ at $T_{\rm{MI}}$ = 3.5 K. The mechanisms for these properties of 1/$T_{2}$ are discussed in Sec. IV B. 
\begin{figure}
\includegraphics[scale= 0.37]{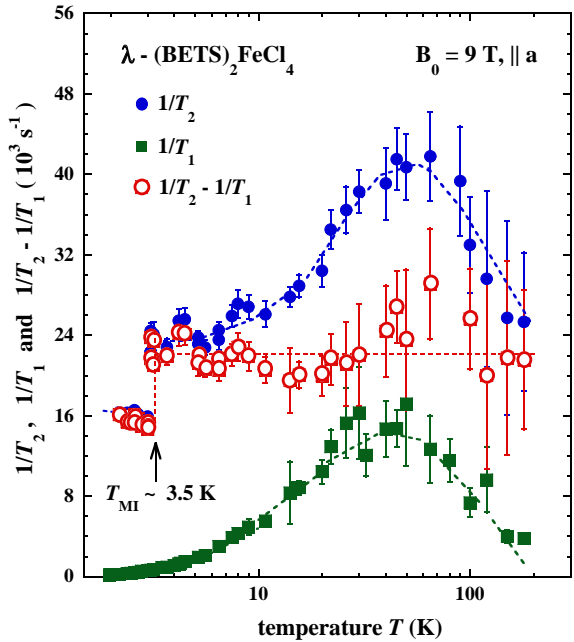}
\caption{(Color online) Proton 1/$T_{2}$ (upper blue curve), 1/$T_{1}$ (lower green curve), and the difference 1/$T_{2}$ $-$ 1/$T_{1}$ (middle red curve) as a function of $T$ for spectral peak 4 in the single crystal of $\rm{\lambda}$-(BETS)$_{2} $FeCl$_{4}$ with $\bf{B}_{0}$ = 8.9885 T $||~a$. The dashed lines are guides to the eye. \label{fig8}}
\end{figure}
\section{discussion and interpretation}
     The main challenge here is to identify and apply the various interactions responsible for the details of the measured $M(2\tau)$. Usually, the main considerations for $M(2\tau)$ are the dipolar interactions among identical and different nuclei, the static and fluctuating magnetic fields from nearby electron moments, and 1/$T_{1}$ of the nuclei under study.\cite{slichter} Another important point is whether the NMR spectrum is strongly inhomogeneously broadened on a microscopic distance scale, which can cause many of the spins of one nuclear species act as ``inequivalent'', or ``detuned'' spins.\cite{ansermet} The latter case applies to our work, where the $\sim$ 8 MHz width of $\chi^{''}(\nu)$ (Fig. 3) from the Fe$^{3+}$ moment is very large in comparison to the homogeneous line width 1/($\pi T_{2}$) $\sim$ 5 kHz (Fig. 4).
\subsection{Origin of the slow beats in $M(2\tau)$}
     An important question is: what is the physical origin of the slow beats observed in $M(2\tau)$? Since the static field from the Fe$^{3+}$ moments provides the large splitting that detunes the protons, it is not possible for this field to directly generate the small value obtined for $f_{B}$. There can be an indirect effect because it determines $\Delta\nu$ for the each proton site, and therefore to which spectral peak it contributes. Also, since the correlation time of the Fe$^{3+}$ moments is expected to be very short because of their large interactions,\cite{wu-lbetsp3} their fluctuations are not expected to determine $f_{B}$.

     A promising mechanism to produce the slow beats in $M(2\tau)$ is the dipole-dipole interaction between the nuclei in the presence of strong inhomogeneous broadening, as reported by Ansermet et al \cite{ansermet} in their work on a CO layer chemisorbed on Pt particles. Below, their model is adapted to the conditions of our measurements. In addition to the inhomogeneous broadening, these conditions are: (1) every proton site ($i$) is occupied and (2) each site has one nearest neighbor from the same CH$_{2}$ unit ($j$ = 1), two other near neighbors from the adjacent CH$_{2}$ group ($j$ = 2, 3) at the same end of the BETS molecule, and many more distant neighbors from other CH$_{2}$ units ($j$ $>$ 3).\cite{wu-lbetsp3}

     Because the inhomogeneous broadening detunes the spins at the fifteen other sites in a unit cell, the probability of mutual spin flips is greatly reduced for calculating $f_{B}$ for a given $i$. Thus, it will be assumed that all proton pairs are detuned. The Hamiltonian ($H_{dij}$) which produces a frequency contribution ($\omega_{ij}$) to the slow beat frequency ($\omega$) from the dipolar coupling between two detuned protons, using the quantization direction $\mathbf{z} \parallel \mathbf{B}_{0}$, is then \cite{ansermet}
\begin{equation}
H_{dij}=\frac{\gamma_{i}\gamma_{j}\hbar^{2}I_{iz}I_{jz}}{r_{ij}^{3}}(1-3\cos^{2}\theta_{ij}),\\
\end{equation}
where $\mathbf{r}_{ij}$ is the vector connecting sites $i$ and $j$, $\theta_{ij}$ is the angle between $\bf{r}_{ij}$ and $\bf{B}_{0}$, $I_{iz}$ and $I_{jz}$ are the $z$-components of the nuclear spin operators ($\bf{I}$), and $\gamma$ is the gyromagnetic ratio of the nuclei (for protons, $I_{iz}$, $I_{jz}$ = $\pm$ 1/2). For proton $i$, the dipole frequency contribution from proton $j$ is
\begin{equation}
\omega_{ij}=\pm\frac{\gamma^{2}\hbar}{2r_{ij}^{3}}\left(1 - 3\cos^{2} \theta_{ij}\right),
\end{equation}
and the total echo amplitude [$m_{i}(2\tau)$] associated with this term for spin $i$ is \cite{ansermet}
\begin{equation}
m_{i}(2\tau)=\cos\left(\sum_{j=1}^{\infty}2\omega_{ij}\tau \right) \simeq \cos\left(\sum_{j=1}^{3}2\omega_{ij}\tau\right).
\end{equation}
This approximation is used for the summation because the near neighbor spins ($j$ $\geq$ 4) have a large number of small terms with alternating signs resulting a negligible effect on $m_{i}(2\tau)$.

     The term $\sum_{j=1}^{3}2\omega_{ij}\tau$ can have both large and small values, depending on which proton site it represents. The reason for this is that the different values of $\theta_{ij}$ for the different nearest neighbor protons can give both larger and smaller values of $\sum_{j=1}^{3}2\omega_{ij}\tau$ (Figs. 2 and 4). A simple way to handle this situation is to interpret $A$ as the fraction of nuclei in the echo for which $\sum_{j=1}^{3}2\omega_{ij}\tau\simeq$ 0 and the term 1 $-$ $A$ as the fraction for the rest with the evolved phase $\sum_{j=1}^{3}2\omega_{ij}\tau = 4\pi f_{B}\tau$. This then gives the ensemble average contribution ($<m(2\tau)>$) 
\begin{equation}
<m(2\tau)> = \left[A+(1-A)\cos(2\pi f_{B}2\tau)\right]
\end{equation}
to $M(2\tau)$. In addition to these terms, there is a multiplicatve factor $\exp(-2\tau/T_{2})$. When these items are combined, one obtains Eq. (1), which provides an excellent fit to the data (Figs. 2 and 5).

     It turns out that a resonably good fit to the echo modulation of peak 4 at 3.2 K can be obtained using the above model and the proton-proton dipolar coupling of site H14 with its three nearest neighbors. Two of the justifications for this approach are: (1) as seen in the bottom of Fig. 3, it is reasonable to consider that proton sites 1, 3, 7, and 14 can contribute to the signal in peak 4, and (2) because the frequency range used for the echo has been narrowed to $\pm$ 60 kHz, which is smaller than the typical calculated shift between these sites, it is expected that a single proton site will dominate the echo signal obtained from the measurements. When one calculates the 8 values of $f_{B}$ associated with the up/down alignment of the 3 nearest neighbors (not shown in detail here) for proton sites 1, 3, 7, and 14, it turns out that the best fit to the data comes from the proton site 14, whose values for $f_{B}$ are $\pm$ 1.02 kHz, $\pm$ 1.57 kHz, $\pm$ 9.39 kHz, and $\pm$ 9.94 kHz, all of which are much smaller than the shifts from the dipolar field of the Fe$^{3+}$.
\begin{figure}
\includegraphics[scale= 0.35]{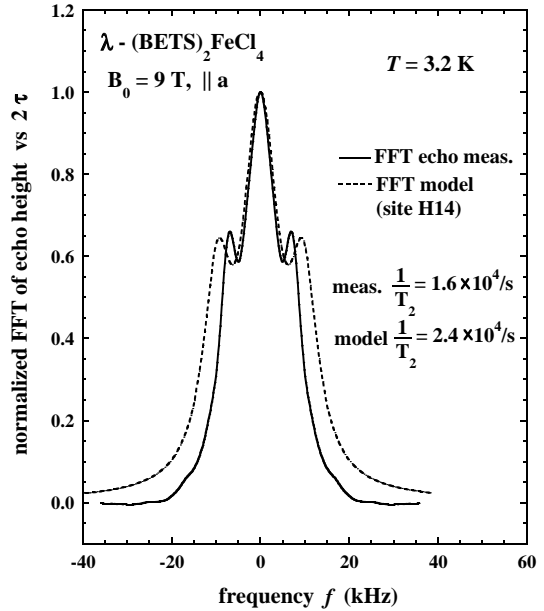}
\caption{Normalized FFT of the measured (solid line) and model (dashed line) $M(2\tau)$ at 3.2 K as a function of $f$ for the single crystal of $\rm{\lambda}$-(BETS)$_{2} $FeCl$_{4}$ with $\bf{B}_{0}$ = 8.9885 T $||~a$. The model value 1/$T_{2}$ = 2.4 $\times$ 10$^{4}$ s$^{-1}$ is used to provide a good fit to the data. \label{fig9}}
\end{figure}

     A comparison of this model to the observed echo decay at 3.2 K in the frequency domain for the peak 4 is shown in Fig. 9. There, normalized FFT of the measured (solid line) and model (dashed line) $M(2\tau)$ at 3.2 K are plotted as a function of the frequency $f$ for $\bf{B}_{0}$ = 9 T $\parallel$ $a$. For this model, the value 1/$T_{2}$ = 2.4$\times$ 10$^{4}$ s$^{-1}$ has been chosen to provide the best fit to the data.

     Several features of our measurements can be discussed in terms of this kind of model. One is that there is a large variation of $f_{B}$ across the spectrum (Figs. 2 and 4). It is attributed to variations in $\omega_{ij}$ (mainly from variations in $\theta_{ij}$) for the different proton sites. Another is the variation in $A$ as the PM-AFI boundary is traversed (Fig. 7). One possible mechanism for that (and the change in 1/$T_{2}$, Fig. 8) is that because of the change in the Fe$^{3+}$ polarization across the PM-AFI phase transition, the contributions of the proton sites responsible for the peak 4 are changed. 
\subsection{The mechanism of the $^{1}$H-NMR spin-echo decay rate 1/$T_{2}$}
     Because of the complexity of the mechanisms for 1/$T_{2}$, it is difficult to give a precise interpretation of the results. In general, the mechanisms that may contribute to the value of 1/$T_{2}$ obtained from the fits to the data are the proton-proton dipolar interactions in a strongly inhomogeneously broadened spectrum, the proton spin-lattice relaxation rate 1/$T_{1}$ (from the power spectrum of the transverse magnetic field fluctuations of the Fe$^{3+}$ moments at $\nu_{0}$), and the magnetic field fluctuations from the Fe$^{3+}$ moments parallel to $\mathbf{B}_{0}$ on the time scale of $T_{2}$. Because of the relatively large spin-spin interactions among the Fe$^{3+}$ moments, which correspond to a fluctuation correlation time $\tau_{c}$ $\sim$ 3 $\times$ 10$^{-9}$ s from the dipole-dipole interaction ($\sim$ 20 G) and $\tau_{c}$ $\sim$ 5 $\times$ 10$^{-12}$ s from the total exchange interaction ($\sim$ 1.7 K),\cite{wu-lbetsp3} it is expected that the field fluctuations from the Fe$^{3+}$ moments parallel to $\mathbf{B}_{0}$ will not contribute significantly to $1/T_{2}$.

     Figure 10 shows the large variation of 1/$T_{2}$, along with the values of $f_{B}$, and 10$\times (1-A)$, plotted as a function of the frequency shift ($\Delta f_{4}$) from the center of the peak 4 at 5.2 K with $\bf{B}_{0}$ = 9 T $\parallel a$. These data were obtained by averaging 10 $-$ 15 points (bandwidth = 122 $-$ 183 kHz) of the FFT of the spin echo signals for each $\Delta f_{4}$ by fits using Eq. (1). The bandwidth of 122 kHz is shown as the typical measured width at the bottom and the vertical error bars are the variance obtained from the fit. Figure 10 indicates that as $\Delta f_{4}$ is varied from left to right, $f_{B}$ increases by about 12 $\%$, $(1-A)$ increases by about 33 $\%$ and 1/$T_{2}$ increases by a factor of 2.17. The latter is about 7.5 times larger than the reduction in 1/$T_{2}$ at the PM-AFI phase boundary (Fig. 8). 

     There are at least two kinds of contributions to 1/$T_{2}$ from the proton dipole-dipole interactions. One is a dynamic interaction from the mutual spin flips between protons that are not detuned. Because of the large inhomogeneous broadening throughout the sample, it is expected that only the protons at the same site,which have the same local field, can participate in this process. This contribution is examined in more detail below.

     The other contribution is static. It comes from the variations in $m_{i}(2\tau)$ for a given site. That leads to a distribution of frequencies in the time domain that corresponds to a broadening of the FFT in the frequency domain. It can include peaks at $f_{B}$ = 0 and finite $f_{B}$ whose width corresponds to the value of 1/$T_{2}$ obtained from the fit of the experimental data with Eqs. (1) and (5).
\begin{figure}
\includegraphics[scale= 0.39]{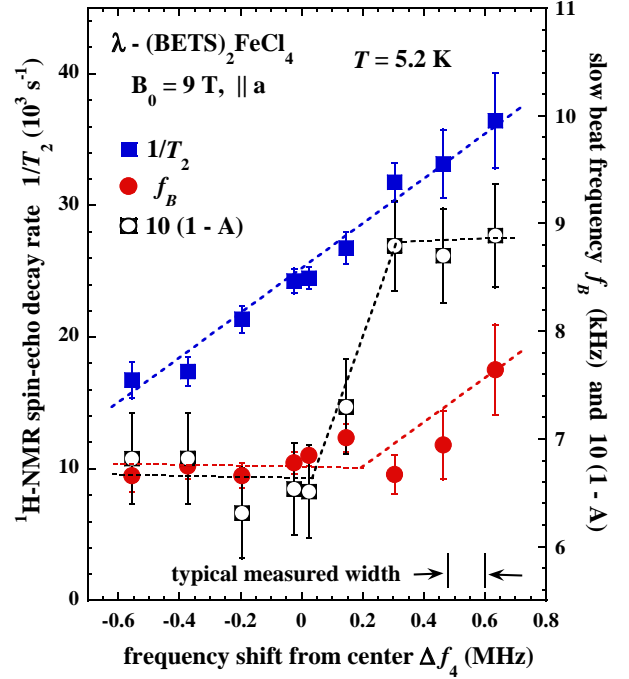}
\caption{(Color online) 1/$T_{2}$, $f_{B}$, and 10$\times(1-A)$ as a function of the frequency shift ($\Delta f_{4}$) from the top of the peak 4 at 5.2 K for the single crystal of $\rm{\lambda}$-(BETS)$_{2} $FeCl$_{4}$ with $\bf{B}_{0}$ = 8.9885 T $||~a$. \label{fig10}}
\end{figure}

     From Fig. 8, it is seen that in the PM phase up to 30 K, 1/$T_{2}$ $-$ 1/$T_{1}$ is constant. At higher $T$ it is constant, but with a large uncertainty. Therefore, we assume that in the PM phase
\begin{equation}
\frac{1}{T_{2}}=C+\frac{1}{T_{1}},\\
\end{equation}
where the constant $C$ $\sim$ 2.3 $\times$ 10$^{4}$ s$^{-1}$. In the AFI phase, the same relation applies, but with $C$ $\sim$ 1.6 $\times$ 10$^{4}$ s$^{-1}$. In many cases, Eq. (6) is a valid representation of experimental results. \cite{slichter}

     Following the discussion above, it is assumed that $C$ = $C_{1}$ + $C_{2}$, where $C_{1}$ is the static contribution from the detuned protons [Eq. (2)] and $C_{2}$ is from the mutual spin flips of protons at the same site on different molecules that remain tuned in the PM phase. In this case, the Hamiltonian ($H_{t,ij}$) is \cite{slichter}
\begin{equation}
H_{t,ij}=-\frac{\gamma_{i}\gamma_{j}\hbar^{2}}{4r_{ij}^{3}}(1-3\cos^{2} \theta_{ij})(I_{i}^{+}I_{j}^{-}+I_{i}^{-}I_{j}^{+}),\\
\end{equation}
where $I^{+}$ and $I^{-}$ are the raising and lowering operators. 

     A calculation of Eq. (7) for one site H14 proton with its five nearest neighbors (details not shown) provides a direct contribution of  $\sim$ 0.7 $\times$ 10$^{3}$ s$^{-1}$ to $C_{2}$, which is very small compared to the measured value $\sim$ 2.3 $\times$ 10$^{4}$ s$^{-1}$. However, it is also expected that there is an indirect dynamic contribution to $C_{2}$ that comes from the mutual spin flips of the near neighbors of a given site, which will also increase its echo decay rate. This contribution is difficult to calculate and is beyond the scope of this paper. Therefore, $\sim$ 0.7 $\times$ 10$^{3}$ s$^{-1}$ is a lower limit for the value of $C_{2}$ for site H14, and the actual value could be substantially larger. Under these circumstances, it is straightforward to obtain $C$ from the experimental results, but it is difficult to separate it into $C_{1}$ and $C_{2}$.

     The next point is how to interpret the discontinuity in $C$ for the peak 4 at the PM-AFI transition, where it drops by $\Delta T_{2}^{-1}$ $\sim$ 7.0 $\times$ 10$^{4}$ s$^{-1}$ ($\sim$ 34 $\%$). It is expected that the overall mechanism for $\Delta$ $T_{2}^{-1}$ is related to the small reorientation of the Fe$^{3+}$ moments that occurs at the transition, but whose details are not presently known. Even though this reorientation should be small because the Zeeman splitting ($\sim$ 12.1 K at 9 T) is large compared to $T_{\rm{MI}}$ and the total interaction between the Fe$^{3+}$ moments ($\sim$ 1 K),\cite{wu-lbetsp3} it could still have a significant effect through changing the local field at each of the proton sites. 

    Under these circumstances, one possible mechanism is that at the transition the spatial period for the dipole field of the Fe$^{3+}$ moments is increased, thereby reducing the density of tuned spins and reducing $C_{2}$. Another possibility follows from the variation of $1/T_{2}$ across the peak 4 (Fig. 10). It is expected that such a change in the Fe$^{3+}$ dipole field will change the distribution of sites that contribute to the part of the peak 4 at which the measurement was made. Since the change of $1/T_{2}$ across the peak 4 is more than a factor of two, the $\sim$ 34 $\%$ drop in $1/T_{2}$ should correspond to a relatively small redistribution of source sites and could be the main mechanism for $\Delta T_{2}^{-1}$. Although these mechanisms are possible ones, they can not be verified within the framework of the present results. It is also possible that a different mechanism is responsible for $\Delta T_{2}^{-1}$.

     In summary, from the measurements of 1/$T_{2}$ over all $T$, it is seen that the proton-proton dipole interactions dominate 1/$T_{2}$ in $\rm{\lambda}$-(BETS)$_{2}$FeCl$_{4}$ below 20 K. At higher $T$, there is a significant contribution from $1/T_{1}$ and it is likely that the proton dipolar part is independent of $T$ throughout the PM phase. There are also large variations in 1/$T_{2}$ as a function of the spectral peak (Fig. 4) and across the peak 4 (Fig. 10). At the PM-AFI transition, measurements on the center of the peak 4 show a $\sim$ 34 $\%$ drop in 1/$T_{2}$. Two possible main mechanisms for this drop are: (1) that the small change in the local field from the reorientation of the Fe$^{3+}$ reduces the number of protons that can undergo mutual spin flips, and (2) that the proton sites responsible for the center of the peak 4 are changed by the change in the local field. The discontinuity in 1/$T_{2}$ and the lack of thermal hysteresis in its change at 3.5 K is evidence that the PM-AFI transition is weakly first order. 
\section{Conclusions}
     NMR measurements of the proton spin echo decay in a small ($\sim$ 4 $\mu$$g$) single crystal of the organic conductor $\rm{\lambda}$-(BETS)$_{2}$FeCl$_{4}$ are reported for $\mathbf{B}_{0}$ = 9 T along the $a$ axis over the range 2.0 K $\leq$ $T\leq$ 180 K. Also, the electrical resistivity as a function of $T$ at several values of $\mathbf{B}_{0}||$ $a$ on the same sample is reported.

     The form of the echo decay is analyzed using a simple model to obtain 1/$T_{2}$ and $f_{B}$ for the NMR spectrum peak 4 as a function of $T$. It is found that below $\sim20$ K, where it can be well identified, $f_{B}$ $\sim$ 7 kHz independent of $T$. Its origin is from the dipole-dipole interaction among the protons, which are strongly detuned by the large, inhomogeneous local magnetic field from the Fe$^{3+}$ moments. It is also found that there is a large variation in 1/$T_{2}$ for the six different spectral peaks and a large variation across the peak 4. 

     In the PM phase, the results for 1/$T_{2}$ show a temperature independent part that is attributed to the dipole-dipole interaction of both tuned and detuned proton pairs and a $T$-dependent part from 1/$T_{1}$. 

     On going from the PM to AFI phase at 3.5 K, there is a discontinuous 34$\%$ drop in 1/$T_{2}$ that is evidence that the transition is first order, in agreement with prior work. \cite{mori, watanabe} Two possible mechanisms for this drop are presented. One involves the reduction of proton pairs that conserve energy in mutual spin flips because of the additional field inhomogeneity caused by the Fe$^{3+}$ realignment at the transition. The other is related to a change in the proton sites that contribute to the echo of the peak 4 caused by the shift in local field at the transition. 
\begin{acknowledgments}
      This work is supported at UCLA by NSF Grant No. DMR-0334869 (W.G.C.) and 0520552 (S.E.B.), partially supported at NHMFL by NSF NO. DMR-0084173 under cooperative agreement, and at Indiana by Petroleum Research fund NO. ACS-PRF 33912-AC1. The authors thank F. Zamborszky and F. Zhang for helpful discussions.
\end{acknowledgments}

\end{document}